\documentclass[aps,prb,twocolumn,amsfonts,superscriptaddress,showpacs]{revtex4}

\usepackage{pst-grad}
\usepackage{amsmath,amssymb,mathrsfs}
\usepackage{latexsym}
\usepackage{graphicx} 
\usepackage{epstopdf}
\usepackage{graphicx,epstopdf,color}
\usepackage{amsfonts}
\usepackage{amsmath,amssymb,mathrsfs}
\usepackage{bm}

\newcommand{\bs}[1]{\boldsymbol{#1}}

\newcommand{\vac}{\left|\,0\,\right\rangle}
\newcommand{\ket}[1]{\left|#1\right\rangle}
\newcommand{\bra}[1]{\left\langle#1\right|}

\newcommand{\nn}{\nonumber}
\newcommand{\pa}{\partial}
\newcommand{\sumk}{\sum_{\bs{k}}}
\newcommand{\up}{\uparrow}
\newcommand{\dw}{\downarrow}
\newcommand{\eps}{\varepsilon}
\newcommand{\pd}{{\phantom{\dag}}}

\def\ie{\emph{i.e.},\ }

\def\ea{\emph{et al.}}

\allowdisplaybreaks[1]
\begin{document}
\title{Quantum Spin Hall Insulators with Interactions and Lattice Anisotropy}
\author{Wei Wu}
\affiliation{Department of Physics, Yale University, New Haven, CT 06520, USA}
\affiliation{Beijing National Laboratory for Condensed Matter Physics, Institute of Physics, Chinese Academy of Sciences, Beijing 100190, China}
\author{Stephan Rachel}
\affiliation{Department of Physics, Yale University, New Haven, CT 06520, USA}
\author{Wu-Ming Liu}
\affiliation{Beijing National Laboratory for Condensed Matter Physics, Institute of Physics, Chinese Academy of Sciences, Beijing 100190, China}
\author{Karyn Le Hur}
\affiliation{Department of Physics, Yale University, New Haven, CT 06520, USA}
 \pagestyle{plain}

\begin{abstract}
We investigate the interplay between spin--orbit coupling and electron--electron interactions on the honeycomb lattice combining the cellular dynamical mean--field theory and its real space extension with analytical approaches. We provide a thorough analysis of the phase diagram and temperature effects at weak spin--orbit coupling. We systematically discuss the stability of the quantum spin Hall phase toward interactions and lattice anisotropy resulting in the plaquette-honeycomb model. We also show the evolution of the helical edge states characteristic of quantum spin Hall insulators as a function of Hubbard interaction and anisotropy. At very weak spin--orbit coupling and intermediate electron--electron interactions, we substantiate the existence of a quantum spin liquid phase.
\end{abstract}

\pacs{71.10.-w, 71.70.Ej, 73.20.At, 71.10.Fd}  


\maketitle

\section{Introduction}
Time--reversal invariant topological insulators\,\cite{ti-reviews} -- bulk insulators with metallic surfaces -- are characterized by a $\mathbb{Z}_2$ invariant\,\cite{kane-mele,moore-07prb121306} and cannot be adiabatically connected to trivial band insulator phases unless the single particle gap closes. While $\mathbb{Z}_2$ topological insulators (TIs) are robust against disorder \cite{disorder}, rigorous and general  results about the fate of TIs in the presence of prominent electron--electron interactions are limited \,\cite{ti+int}. Strongly correlated TIs as well as exotic time--reversal invariant Mott insulator phases have been predicted\,\cite{pesin-10np376,young-08prb125316,rachel-10prb075106,goryo-11jpsj044707,raghu-08prl156401,witczak-krempa-10prb165122,kargarian-11prb165112} apart from more conventional magnetically ordered phases. 

By analogy to the quantum Hall effect two--dimensional TIs are also named quantum spin Hall (QSH) insulators. They were originally proposed to be realized in graphene\,\cite{kane-mele} and later also in HgTe/CdTe quantum wells\,\cite{bernevig-06s1757} where subsequent experiments\,\cite{koenig-07s766} measuring a quantized conductance established the field of TIs. 
They possess an odd number of pairs of time-reversal conjugate counter-propagating edge states (the helical edge states)\,\cite{kane-mele,bernevig-06s1757,koenig-07s766,wu-06prl106401}. There are other promising proposals to stabilize the QSH effect in real materials: graphene endowed with heavy adatoms like indium and thallium\,\cite{weeks-11arXiv1104.3282} and  synthesized silicene\,\cite{liu-11arXiv1104.1290} was shown to exhibit a stable QSH phase. Particularly interesting are monolayers or thin films of Iridium--based materials X$_2$IrO$_3$ (X=Na or Li) which have been debated to possibly host QSH phase\,\cite{shitade-09prl256403,singh-10arXiv1006.0437}
since both spin orbit coupling (SOC) and electron--electron interactions are quite strong in such materials. All these systems have in common the underlying honeycomb lattice where recently a gapped quantum spin liquid for intermediate interactions was found using quantum Monte Carlo (QMC)\,\cite{meng-10n847,hohenadler-11prl100403,Wu2}.

Very recent progress within ultracold atoms in tunable optical lattices establishes a second class of systems (beside Iridium based materials) where topological interacting phases might be realized. This progress is two--fold: (i) tunable hexagonal lattices have been realized\,\cite{hexagonal-OL} and (ii) different types of spin--orbit interactions are feasible now\,\cite{SOC-OL}.
Additional electron--electron onsite interactions are considered as a standard tool in optical lattices\,\cite{bloch-08rmp885}. All these achievements in such a  rapidly evolving field indicate the demonstration of topological interating phases within cold atoms in the very near future.

\begin{figure}[b!]
\begin{center}
\includegraphics[scale=0.6]{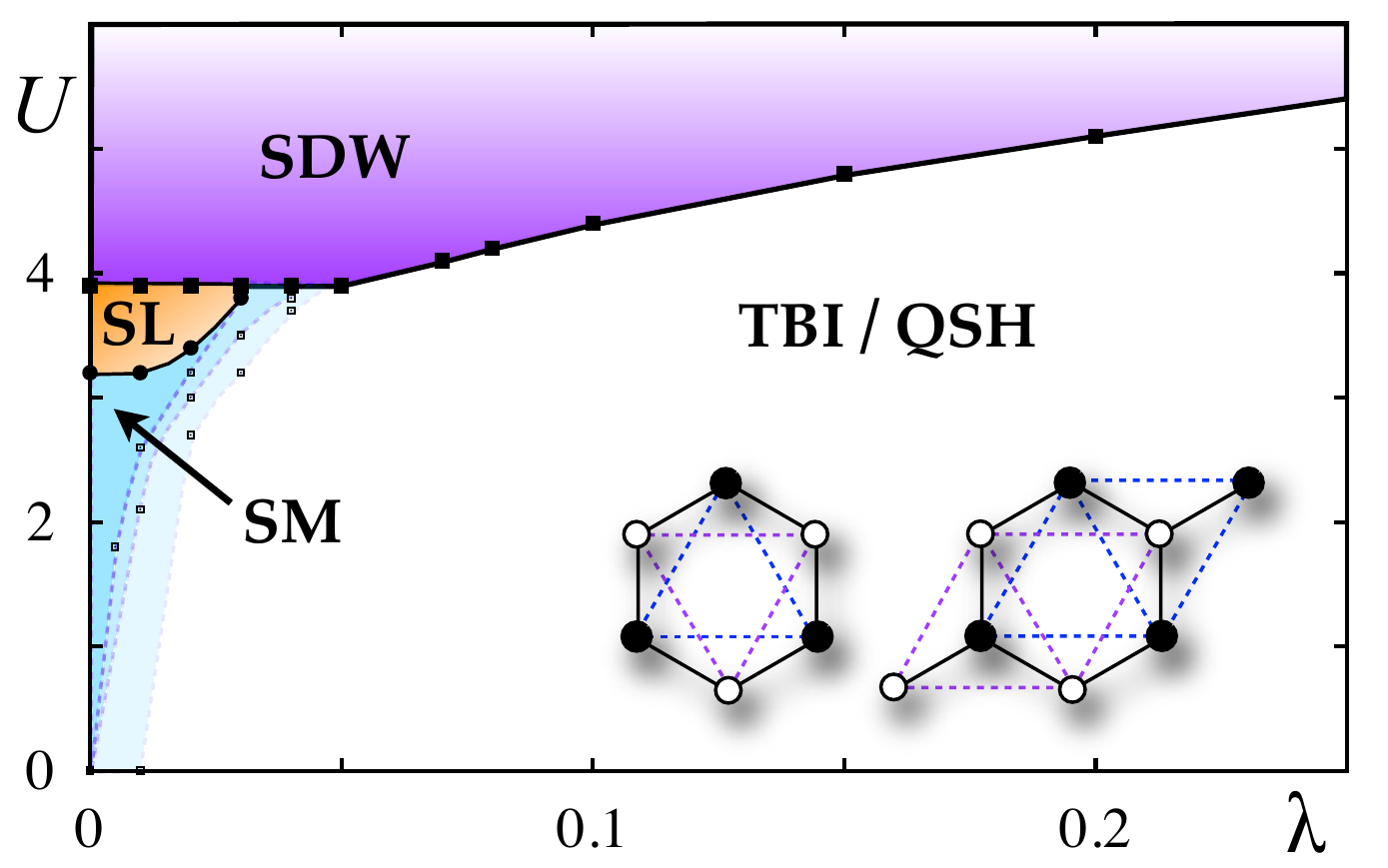}
\caption{(color online). Phase diagram of KMH model within CDMFT, including the four phases: (i) topological band insulator (TBI), (ii) magnetically ordered spin density wave phase (SDW), (iii) non--magnetic insulator phase (SL), and (iv) semi--metal (SM) region which is shown (from right to left) for temperatures $T=0.025$, $0.0125$, and $0.005$. Extrapolating $T\to 0$, the SM region shrinks to a line, see main text. All other phase boundaries are extrapolated to $T=0$.
Inset: Typical clusters as used within CDMFT.}
\label{fig:phase}
\end{center}
\end{figure}

In this paper, we investigate the interplay between SOC and electron--electron interactions on the honeycomb lattice and combine two very paradigmatic models: to capture the non--trivial band topology we consider the Kane--Mele (KM) model\,\cite{kane-mele} without Rashba term and to describe interaction effects the Hubbard model, merging to the Kane--Mele--Hubbard model (KMH)\,\cite{rachel-10prb075106,hohenadler-11prl100403,soriano10,Wu2,yamaji-10arXiv1012.2637,yu-11arXiv1101.0911}. Our goal is to combine the cellular dynamical mean--field theory (CDMFT)\,\cite{kotliar-06rmp865,maier-05rmp1027,wu-10prb245102} and its real space extension with analytical approaches, present our phase diagram at half filling including temperature effects, and thoroughly address the fate of helical edge states as a function of interactions and SOC coupling. Additionally, we introduce the plaquette-honeycomb model which illustrates that the QSH phase is also stable toward lattice anisotropy.

\section{Kane--Mele--Hubbard model}
The KMH Hamiltonian on the honeycomb lattice reads
\begin{equation}\label{ham}
H \!=\! -t \sum_{\langle ij \rangle\sigma} c_{i\sigma}^\dag c_{j\sigma}^\pd + \,i\lambda\!\!\!\sum_{\ll ij \gg\alpha\beta} 
\!\!\!\nu_{ij} c_{i\alpha}^\dag \sigma^z_{\alpha\beta} c_{j\beta}^\pd + U\sum_i n_{i\up} n_{i\dw}
\end{equation}
where $i,j,$ label the sites on the honeycomb lattice, $c_{i\sigma}$ is the electron annhilation operator, $n_{i\sigma}=c_{i\sigma}^\dag c_{i\sigma}^\pd$, $t$ is the hopping 
integral ($t$ is our reference energy scale, and hence we set $t\equiv 1$), $\lambda$ the SOC, $U$ the onsite interaction, and $\nu_{ij}={\rm sgn}[(\hat{\bs{d}}_i\times\hat{\bs{d}}_j)_z]$ where $\hat{\bs{d}}_{i/j}$  are the two vectors along the links from $j$ to $i$ $(\nu_{ij}=\pm 1)$ \cite{kane-mele}. While the Hubbard model respects SU(2) spin  and $C_6$ lattice symmetry, the SOC breaks the spin symmetry down to U(1) and the lattice to $C_3$ while it leaves time reversal symmetry invariant. 
Details of the used CDMFT method and its real space extension are provided in Appendix A.

\begin{figure}[t!]
\centering
\includegraphics[scale=0.42]{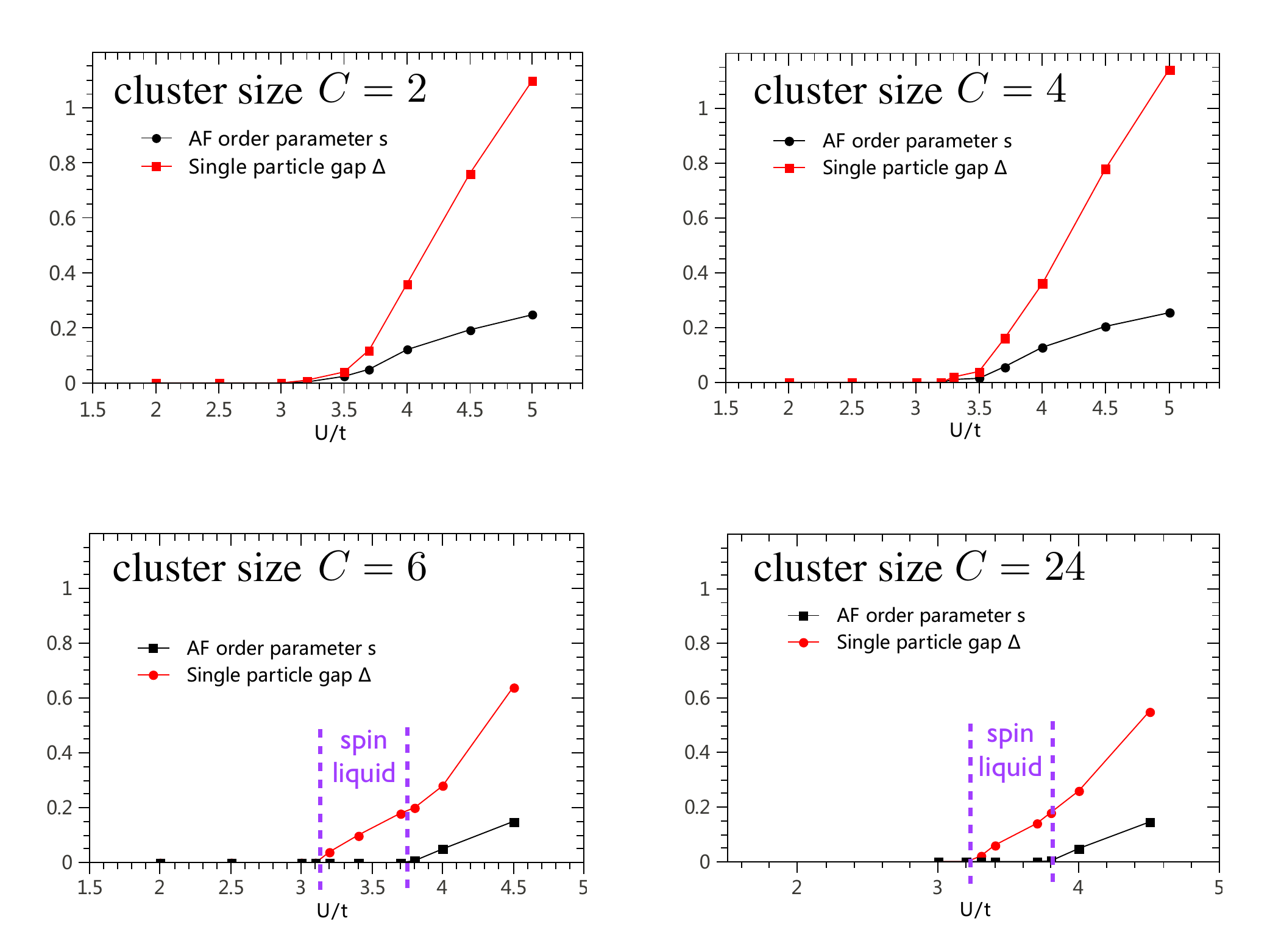}
\caption{(color online). Single particle gap $\Delta_{\rm sp}$ and magnetization $m$ as a function of $U$ for $\lambda=0$. The various panels correspond to different cluster sizes used by the CDMFT method. A minimal cluster size $C=6$ is required to find the non--magnetic insulator phase.}
\label{fig:cluster-size}
\end{figure}

For large interactions, the phase is an easy-plane antiferromagnet\,\cite{rachel-10prb075106}. First, it is useful to start with a mean--field (MF) consideration of the TBI--SDW transition which can also be described within slave--rotor theory\,\cite{rachel-10prb075106} where the condensation of magnetic monopoles results in an $XY$--instability. 
At $\lambda=0$, Eq.\,\eqref{ham} is spin rotationally invariant and a staggered magnetization pointing in any direction will be a good order parameter. For finite $\lambda$, one can learn from the effective spin model\,\cite{spinmodel} (\ie the strong coupling limit of Eq.\,\eqref{ham}, cf.\,Ref.\,\onlinecite{rachel-10prb075106}) that the magnetization lies in the $XY$ plane. Hence we chose the order parameter accordingly, $m= \langle S_i^+\rangle=\langle S_i^x\rangle +i\langle S_i^y\rangle$. From the Hubbard interaction we obtain
$n_{i\up}n_{i\dw}= -S_i^+ S_i^-+(n_{i\up}+n_{i\dw})/2$. A standard MF decomposition results in $H_I \approx \sumk U\left( m (b_{\bs{k}\dw}^\dag b_{\bs{k}\up}^\pd  - \,a_{\bs{k}\dw}^\dag a_{\bs{k}\up}^\pd) + m^\star (b_{\bs{k}\up}^\dag b_{\bs{k}\dw}^\pd- a_{\bs{k}\up}^\dag a_{\bs{k}\dw}^\pd) \right) + {\rm const.}$ where the operators $a_{\bs{k}\sigma}$ and $b_{\bs{k}\sigma}$ are associated with the two sublattices. 
Eventually one obtains the MF spectrum $\eps_{\pm}^{\rm MF} = \pm \sqrt{ |g(\bs{k})|^2 + \gamma(\bs{k})^2 + U^2|m|^2} +{\rm const.}$ and from there we find the MF equation. Note that $g(\bs{k})$ is the nearest neighbor hopping and $\gamma(\bs{k})$ the second neighbor (spin--orbit) Haldane term\,\cite{haldane88prl2015}. The transition line differs slightly from the analog calculation with the order parameter $\langle S_i^z\rangle$. Mainly, the form of $\eps_{\pm}^{\rm MF}$ reveals that the mean field approximation causes another mass term which does not compete with the SOC mass $\gamma(\bs{k})$. Consequently, when passing from the TBI to the SDW phase we do not expect closing of the single particle gap. Indeed, within CDMFT no closing of the single particle gap is observed (while the gap has a local minimum at the transition). 
We also computed $\langle S_i^x \rangle$ which becomes finite at the TBI-SDW transition (see also App.\,A); note that this transition is of 3D XY universality at $T=0$.

\begin{figure}[t!]
\begin{center}
\includegraphics[scale=0.58]{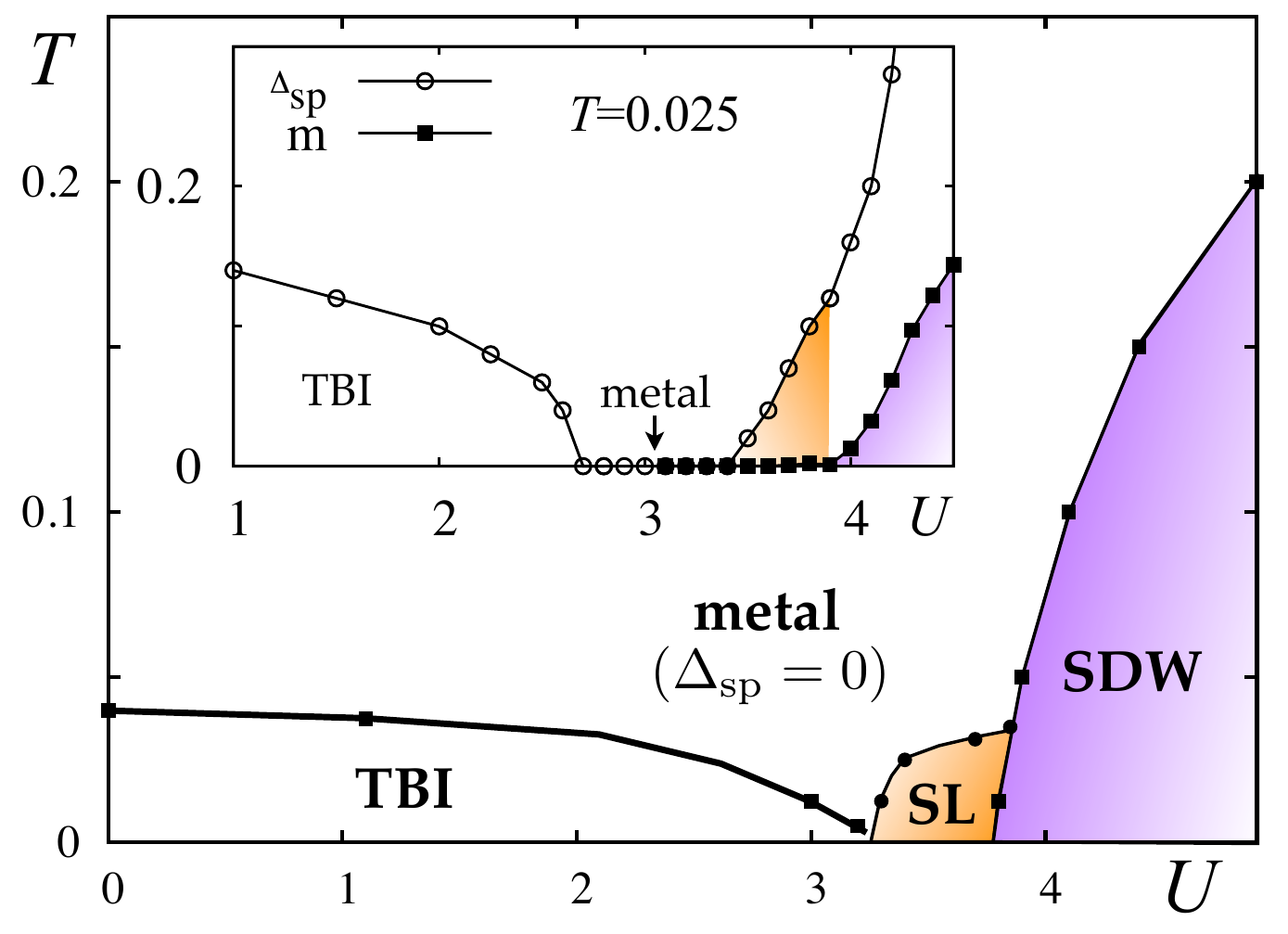}
\caption{(color online). Temperature dependence of the phase diagram at SOC $\lambda=0.02$. Inset: Single-particle gap $\Delta_{\rm sp}$ and magnetization $m$ vs.\ $U$ is shown for $\lambda=0.02$ and $T=0.025$.}
\label{fig:temp}
\end{center}
\end{figure}

For very weak SOC we identify a small phase at intermediate $U$ which is non--magnetic, exhibits a finite spin gap, and is separated from the TBI phase by closing of the single--particle gap (see Fig.\,\ref{fig:temp}). This phase 
is reminiscent of the recently found quantum spin liquid (SL) phase\,\cite{meng-10n847,hohenadler-11prl100403}, in particular, position and shape of this non--magnetic insulator phase essentially coincides with the SL phase found within QMC\,\cite{hohenadler-11prl100403}.
Here, we shall mention that since the CDMFT method approximates the self--energy by restricting it to the chosen cluster (see inset of Fig.\,\ref{fig:phase}), the correlation length of this SL phase cannot be inferred. Note that at least a six--site cluster is required to observe the SL phase, see Fig.\,\ref{fig:cluster-size}. 
We have plotted single particle gap $\Delta_{\rm sp}$ and magnetization $m$ for $\lambda=0$ and various cluster sizes $C$. For $C=2$ and $C=4$ single particle gap and magnetization become finite simultaneously while for $C\geq 6$ the onset of both quantities occurs for different $U$ indicating the SL phase. The existence of such possible spin-gapped phases on the honeycomb lattice has also been addressed theoretically\,\cite{sl-theory}.

\begin{figure}[t!]
\centering
\includegraphics[scale=0.6]{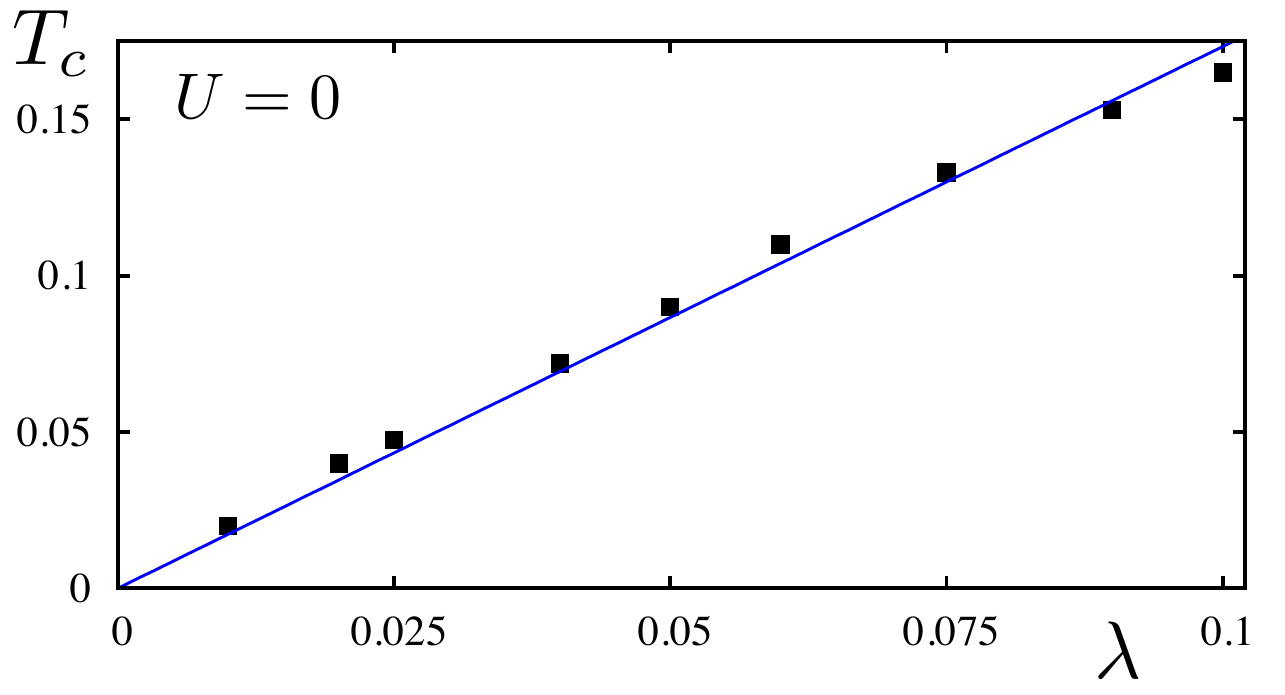}
\caption{$T_c$ as function of $\lambda$ for TBI phase at $U=0$. The blue line is a linear fit.}
\label{fig:tc-tbi}
\end{figure}

In the phase diagram of Fig.\,\ref{fig:phase} there is an additional semi--metal (SM) phase where the Dirac cones of graphene are preserved. Using QMC as impurity solver, we systematically discuss temperature effects at weak SOC. We observe that the SL phase appears at very low temperatures (see Fig. 2) and that the SM domain prominently develops with temperature as the TBI gap
is also very small when $\lambda\rightarrow 0$. For $T\to 0$, the SM phase reduces to a line pointing along the $\lambda=0$ axis which survives until relatively large interactions. From Renormalization Group point of view, this SM line at $\lambda=0$ is known to be stable for weak $U$\,\cite{kotov-10arXiv:1012.3484}.  Hence, we show the SM region for $T=0.025$, $0.0125$, and $0.005$ in Fig.\,\ref{fig:phase}. Remember that both the SL and TBI phases at $\lambda \leq 0.03$ possess a small single-particle gap. In the inset of Fig.\,\ref{fig:temp} we present the single particle gap and magnetization for $\lambda=0.02$ and $T=0.025$. The evolution of the phase diagram with temperature is shown in Fig.\,\ref{fig:temp} for $\lambda=0.02$.  

We are using continuous time quantum Monte Carlo (CTQMC) as an impurity solver which enables us to access finite temperatures. This is clearly an advantage of the method since experiments are performed at finite temperatures but a careful interpretation of results is still needed. In Fig.\,\ref{fig:temp}, for instance, the SDW phase at finite $T$ violating Mermin-Wagner theorem is clearly an artifact of the CDMFT method. In contrast, the stability of the TBI phase with increasing temperature is reliable and important for experimentalists. From an experimental perspective, it may be interesting to know how $T_c$ can be raised and for what parameter settings $T_c$ is conveniently large. Since the spin orbit gap for $\lambda=0.02$ is tiny, $T_c<0.05$. Increasing $\lambda$ (and the bulk gap), however, leads to an (approximately linear) increase of $T_c$ up to $T_c>0.16$ for $\lambda=0.1$ and $U=0$; see Fig.\,\ref{fig:tc-tbi} where the line is a linear fit. 
We corroborate that the critical temperature $T_c$ follows the gap at $T=0$.
For finite values of $U$ a similar behavior of $T_c$ is obtained. Note that $T_c$ does not necessarily indicate a phase transition from TBI to metal;
to clarify the precise nature of this boundary would require further work.

\begin{figure}[t!]
\begin{center}
\includegraphics[scale=0.6]{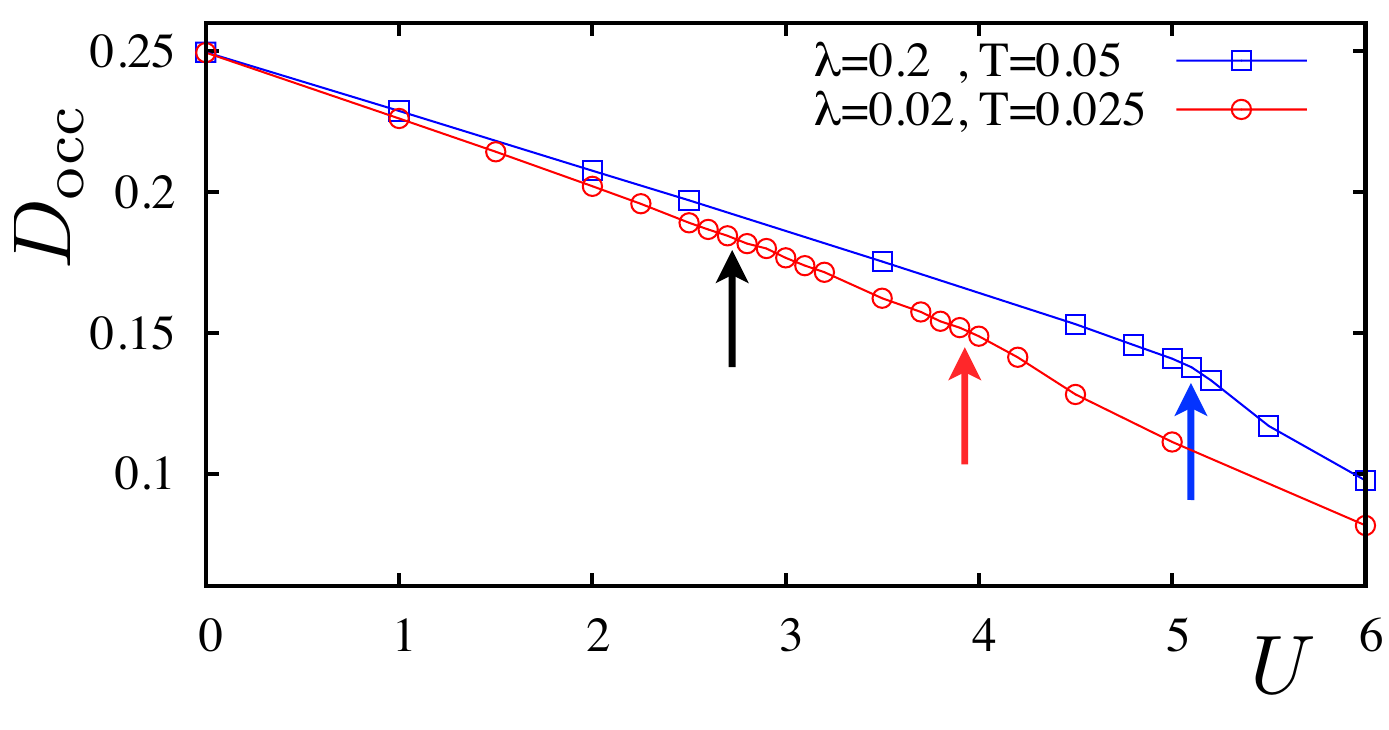}
\caption{(color online). Double occupancy $\langle n_{i\up} n_{i\dw}\rangle$ as a function of $U$ is shown for $\lambda=0.02$ (red circles) and $0.2$ (blue squares). The blue arrow (for $\lambda=0.2$) at $U=5.1$ and the red arrow (for $\lambda=0.02$) at $U=3.9$ mark the phase transition into the magnetically ordered phase. Note, the transition TBI (black arrow) to SM cannot be resolved.}
\label{fig:docc}
\end{center}
\end{figure}

Within CDMFT, we also access the double occupancy $D_{\rm occ} = \langle n_{i\up}n_{i\dw}\rangle = {\pa F}/{\pa U}$ where $F$ is the free energy.  In Fig.\,\ref{fig:docc} we show that $D_{\rm occ}$ which measures the itinerant properties of electrons
is an insightful quantity to detect the magnetically ordered phase\,\cite{wu-10prb245102,gorelik}.
The double occupancy is bounded as $1/4 \geq D_{\rm occ} \geq 0$. For $U=0$, the ground state is a TBI with occupied lower and empty upper bands,
$\ket{\rm TBI}=\prod_{\bs{k}} l_{\bs{k}\up}^\dag l_{\bs{k}\dw}^\dag\vac$,
 where $(a_{\bs{k}\sigma}, b_{\bs{k}\sigma})^T = T_\sigma (l_{\bs{k}\sigma}, u_{\bs{k}\sigma})^T$ and the matrices $T_\sigma$ are given by
\begin{equation}
T_\up = \left( \begin{array}{cc} -\alpha_-& -\alpha_+ \\ \beta_- & \beta_+ \end{array}\right),\qquad
T_\dw = \left( \begin{array}{cc} \alpha_+ & \alpha_- \\ \beta_+ & \beta_-  \end{array}\right).
\end{equation} 
with
\begin{eqnarray}
\alpha_\pm&=&\alpha_\pm(\bs{k})=\mathcal{N}_\pm g(\bs{k})[\gamma(\bs{k})\pm\eps(\bs{k})]/|g(\bs{k})|^2\ ,\\ \nn
\beta_\pm&=&\beta_\pm(\bs{k})=\mathcal{N}_\pm\ ,\\ \nn
\mathcal{N}_\pm&=&|g(\bs{k})|/\sqrt{|g(\bs{k})|^2 + [ \gamma(\bs{k}) \pm \eps(\bs{k})]^2}\ .
\end{eqnarray}
Here $g(\bs{k})$ is the nearest neighbor hopping term, $\gamma(\bs{k})$ is the Haldane term (second neighbor spin orbit hopping), and $\eps(\bs{k})=\sqrt{|g(\bs{k})|^2+\gamma(\bs{k})^2}$ is the single--particle Kane--Mele energy spectrum. Essentially, the matrices $T_\sigma$ contain the eigenvectors of the single--particle eigenstates.
The Fermi level lies in the gap, $\eps_F=0$. Choosing a site $i$ that belongs to the sublattice ``a'' we obtain,
\begin{eqnarray}
\nn&&\bra{\rm TBI}   n_{i\up}n_{i\dw} \ket{\rm TBI}= \\[5pt]
\nn&=&\ \frac{1}{N^2} \sum_{\bs{k}_1\bs{k}_2\bs{k}_3\bs{k}_4} e^{-i(\bs{k}_1-\bs{k}_2+\bs{k}_3-\bs{k}_4)\bs{R}_i} \langle  a_{\bs{k}_1\up}^\dag a_{\bs{k}_2\up}^{\phantom{\dag}} a_{\bs{k}_3\dw}^\dag a_{\bs{k}_4\dw}^{\phantom{\dag}} \rangle\\[5pt]
\nn&=&\ \frac{1}{N^2} \sum_{\bs{k}_1\bs{k}_2\bs{k}_3\bs{k}_4} e^{-i(\bs{k}_1-\bs{k}_2+\bs{k}_3-\bs{k}_4)\bs{R}_i}  \\
\nn&&\quad \times\alpha_-^\star(\bs{k}_1)\alpha_-(\bs{k}_2)\alpha_+^\star(\bs{k}_3)\alpha_+(\bs{k}_4) \langle \, l_{\bs{k}_1\up}^\dag l_{\bs{k}_2\up}^{\phantom{\dag}} l_{\bs{k}_3\dw}^\dag l_{\bs{k}_4\dw}^{\phantom{\dag}} \, \rangle \\[5pt]
&=&\ \frac{1}{N^2} \sum_{\bs{k}_1\bs{k}_3} |\alpha_-(\bs{k}_1)|^2 |\alpha_+(\bs{k}_3)|^2\ ,
\end{eqnarray}
where we used in the second last line $\langle \, l_{\bs{k}_1\up}^\dag l_{\bs{k}_2\up}^{\phantom{\dag}} l_{\bs{k}_3\dw}^\dag l_{\bs{k}_4\dw}^{\phantom{\dag}} \, \rangle=\delta_{\bs{k}_1\bs{k}_2}\delta_{\bs{k}_3\bs{k}_4}$ as well as $\langle u_{\bs{k}\sigma}^\dag u_{\bs{k}\sigma}^{\phantom{\dag}}\rangle=0$, $\langle u_{\bs{k}\sigma}^\dag l_{\bs{k}\sigma}^{\phantom{\dag}}\rangle=0$ and $\langle l_{\bs{k}\sigma}^\dag l_{\bs{k}\sigma}^{\phantom{\dag}}\rangle=1$. $N$ refers to the number of unit cells.
Using the relations\,
\begin{equation}
\sumk |\alpha_\pm(\bs{k})|^2 = \sumk |\beta_\pm(\bs{k})|^2 = N/2
\end{equation}
we confirm that
\begin{equation}
D_{\rm occ}\Big|_{U=0} = \bra{\rm TBI} n_{i\up}n_{i\dw}\ket{\rm TBI} = \frac{1}{4}
\end{equation}
independent of $\lambda$ in agreement with CDMFT, see Fig.\,\ref{fig:docc}. 
Above we considered without loss of generality $n_{i\sigma}=a_{i\sigma}^\dag a_{i\sigma}^\pd$. Identical calculations performed with the $b_{i\sigma}$ operators which belong to the other sublattice yields of course the same result.
In the opposite limit, $U\to\infty$, we clearly find $D_{\rm occ}=0$ as a fingerprint of Mott physics since we impose half filling.

\section{Helical edge states}
To describe the edge states associated with QSH insulators, first we apply the concept of the helical Luttinger liquid (HLL)\,\cite{wu-06prl106401}. Hence, we linearize the spectrum around the Fermi points and switch from lattice operators to field operators $\psi_{R\up}(x)$ and $\psi_{L\dw}(x)$ which are right-- and left--moving fields, respectively; we obtain for the non-interacting part
$H_0 = v_F \int dx \left( \psi_{R\up}^\dag i \pa_x \psi_{R\up}^{\phantom{\dag}} - \psi_{L\dw}^\dag i \pa_x \psi_{L\dw}^{\phantom{\dag}}\right)
$. Note that a standard single-particle (disorder) backscattering term $\psi_{R\up}^\dag \psi_{L\dw}^{\phantom{\dag}} + {\rm h.c.}$, which opens up a mass gap in the spinless Luttinger liquid, is not allowed since the model is odd under time-reversal symmetry, $\mathcal{T}^2=-1$. Only two time--reversal invariant interactions are allowed: the forward scattering as well as the umklapp scattering $\sim \psi_{R\up}^\dag \psi_{R\up}^\dag\psi_{L\dw}^\pd\psi_{L\dw}^\pd$ which is not intrinsically present in the Hubbard model. Instead we shall include the forward interaction
$H_I = U \int dx \left( \psi_{R\up}^\dag\,\psi_{R\up}\psi_{L\dw}^\dag\, \psi_{L\dw}^\pd \right)$.
As long as there is no magnetic order in the bulk, $(H_0+H_I)$ can be solved exactly resorting to bosonization which results in power-law decaying spin and charge correlations in the HLL.
This result is also obtained through a spin wave analysis at the edges for weak interactions\,\cite{lee11arXiv:1105.4900}. 
It is worth mentioning
that, in contrast, spin-spin correlation functions in the bulk decay very rapidly\,\cite{rachel-10prb075106}.

At the SDW transition, $H_I$ turns into $H_I\approx -U m\int dx\, \psi_{L\dw}^\dag\psi_{R\up}^\pd + {\rm h.c.}$ with $m=\langle \psi_{R\up}^\dag \psi_{L\dw}^\pd\rangle$; applying the bosonization procedure
and introducing 
the Luttinger parameter $K$ as usual,
the Hamiltonian becomes:
\begin{equation}
H =\! \int dx \frac{v}{2} \left[ \frac{1}{K} \left(\pa_x \phi\right)^2 + K\left( \pa_x \theta \right)^2 \right]
- \frac{U m  \sin{ \sqrt{4\pi}\phi}}{(\pi a)^2}.
\end{equation}
Here, $v$
is the renormalized plasmon velocity and $a$ the lattice spacing, the field $\phi$ contains both spin and charge degrees of freedom and $\pa_x \theta$ and $\phi$ are conjugate variables \cite{wu-06prl106401}. 
The Sine-Gordon term is a relevant perturbation for repulsive interactions (since
$K\ll 2$) and hence, through the pinning of the $\phi$--field, the edge modes acquire a charge
gap at the SDW transition. We can also check that the spin sector at the edges
is described by an Ising order characterized by $\langle S_i^x \rangle \neq 0$ and $\langle S_i^y \rangle = 0$.
This shows that the SDW transition affects the charge sector of the HLL in a non-trivial way resulting in a metal-insulator transition of Kosterlitz-Thouless type. Note that the disordering-ordering transition in the spin sector also influences the charge properties of the edges.
\begin{figure}[t!]
\begin{center}
\includegraphics[scale=0.65]{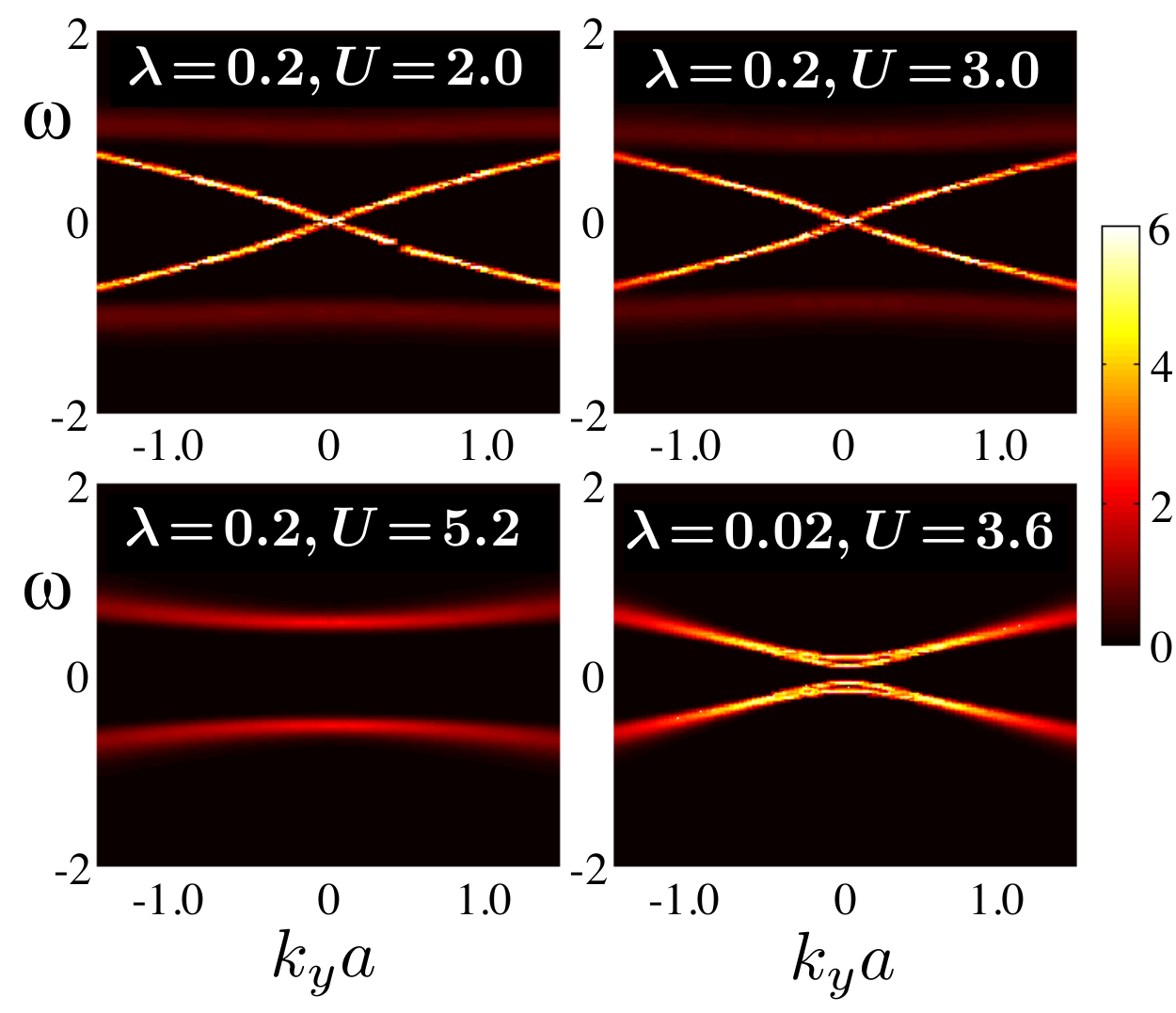}
\caption{(color online). Spectral function $A(k_y,\omega)$ for a cylinder geometry with armchair edges ($L=96$, $T=0.05$). Top: helical edge states of the QSH phase with $\lambda=0.2$ and $U=2.0$ or $3.0$, respectively. Bottom, left: for $U=5.2>U_c$ (SDW phase) the helical edge states disappear while the single particle gap remains finite. Bottom, right: $A(k_y,\omega)$ inside the SL phase: no edge states cross the gap. The color bar corresponds to the intensity of $A(k_y,\omega)$.}
\label{fig:akw}
\end{center}
\end{figure}

In addition, we performed CDMFT simulations on honeycomb cylinders with length $x=(3a/4) L$ with both armchair and zigzag edges ($L=$ number of sites in $x$--direction). By computing the spectral function $A(k_y,\omega)$ we extract the edge state spectrum in the presence of interactions. In contrast to previous work\,\cite{hohenadler-11prl100403} we treat the full microscopic Hamiltonian. We observe the following: (i) The plasmon velocity $v$ associated with the edge modes slightly decreases for increasing $U$ as expected from the HLL.
(ii) The intensity of the spectral function decreases with increasing $U$. (iii) The edge modes gap out when 
$\langle S_i^x\rangle$ becomes finite. In Fig.\,\ref{fig:akw} we show exemplarily the edge modes for fixed $\lambda=0.2$, armchair boundary conditions, and $U=2.0$, $3.0$, and $5.2$.

\begin{figure}[t!]
\begin{center}
\includegraphics[scale=0.6]{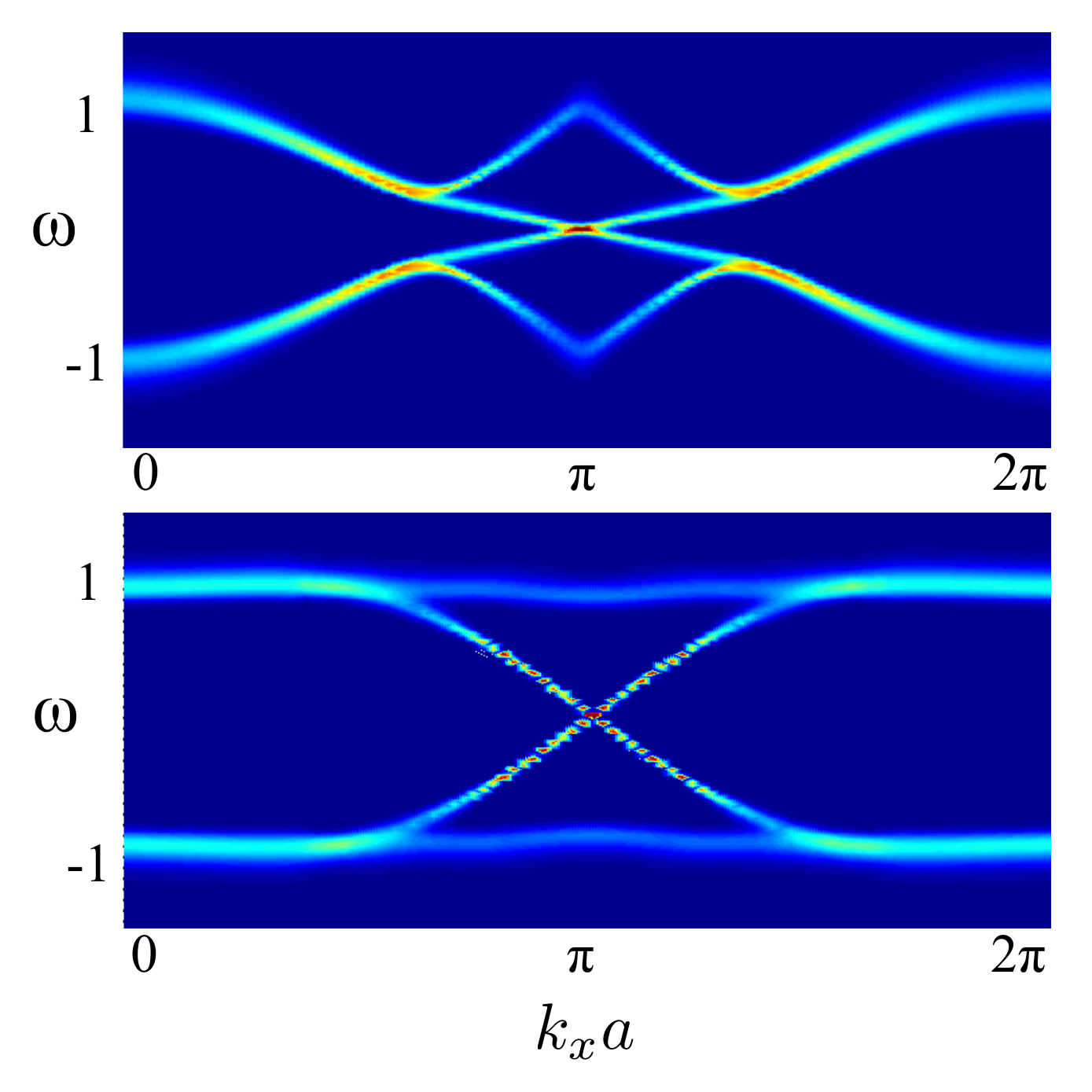}
\caption{(color online). Spectral functions for a zigzag-ribbon at $\lambda=0.05$ and $U=2.0$ (top) and at $\lambda=0.2$ and $U=2.0$ (bottom) are shown.}
\label{fig:zigzag}
\end{center}
\end{figure}

We further performed computations for the spectral function on zigzag--ribbons.
Now, in contrast to the armchair--case, the periodic boundary conditions are imposed in $x$--direction yielding $k_x$ as the good quantum number.
For zigzag edges one extracts the more familiar spectrum\,\cite{kane-mele} of the KM model at small $\lambda$ (see Fig.\,\ref{fig:zigzag}, top): the edge states connect the upper band of a gapped Dirac cone $\bs{K}$ with the lower band of the 
other gapped Dirac cone $\bs{K}'$ and vice versa. 
In addition, we have also shown the situation for stronger SOC $\lambda=0.2$ (see Fig.\,\ref{fig:zigzag}, bottom) where the bulk spectrum is relatively flat. Both spectral functions are computed for $U=2.0$.
The main findings (i) -- (iii) obtained for armchair--ribbons also apply for zigzag--ribbons.

We also computed the spectral function on a cylinder for parameters $U$ and $\lambda$ which belong to the spin liquid phase (see Fig.\,\ref{fig:akw}). While the single particle gap is very small we did not find edge states inside the gap. While it is not known whether the {\it true} SL exhibits edge states, the non-magnetic insulator phase found within CDMFT is topologically trivial and absence of edge states expected.

\begin{figure}[t!]
\centering
\includegraphics[scale=0.6]{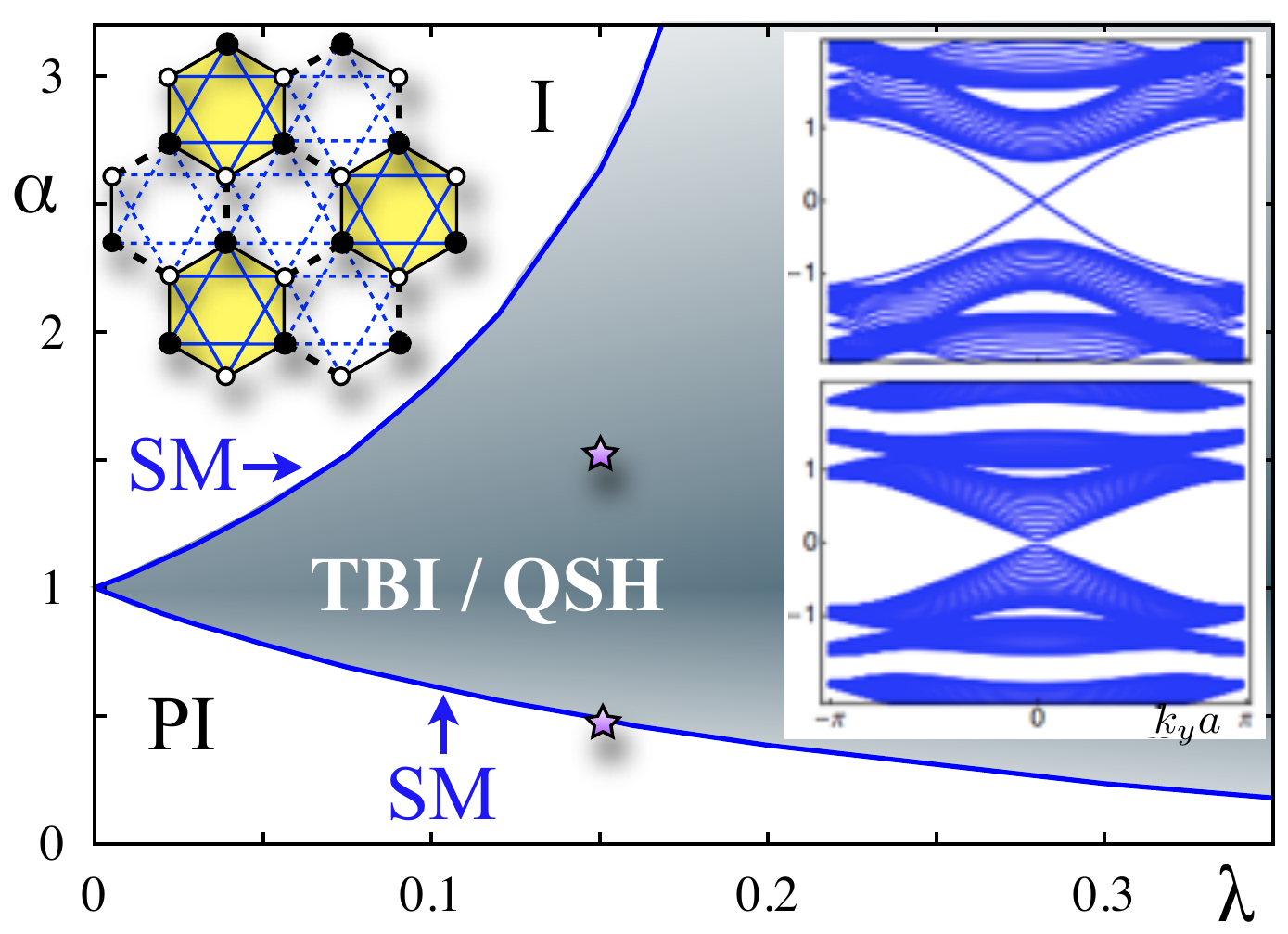}
\caption{(color online). Zero-temperature $\alpha$--$\lambda$ phase diagram of the plaquette honeycomb model at $U=0$. The $\alpha=1$ line corresponds to the KM model. Spectra for armchair--ribbons ($L=96$) are shown at $\lambda=0.15$, $\alpha=1.5$ (top, QSH phase) and $\alpha=0.48$ (bottom, entrance of PI phase). Blue lines correspond to SM.}
\label{fig:aniso}
\end{figure}

\section{Plaquette honeycomb model}
Next, we show that the QSH phase is not only stable towards relatively large interactions  but also to lattice anisotropy resulting in the plaquette-honeycomb model, as sketched in the inset of Fig.\,\ref{fig:aniso}. Hopping amplitudes which connect yellow plaquettes are $t'=\alpha t$ while hopping amplitudes inside a cluster are $t$. Spin--orbit couplings $\lambda$ and $\lambda'=\alpha\lambda$ are analogously defined. In the absence of spin-orbit couplings, note that this type of anisotropy immediately opens a gap at the Dirac points. Mostly when $\alpha\rightarrow 0$, we check that the spectrum exhibits flat bands associated with localized states on a given hexagon, hence resulting in the plaquette insulator (PI) in Fig.\,\ref{fig:aniso}. Also for large $\alpha$, the system is an insulator (I) at the Fermi level. For finite SOC 
$\lambda$, we find that the QSH phase is stable over a large region of the phase diagram (containing the KM model at $\alpha=1$). The gap closes at the phase transition between the PI/I and the QSH phase yielding a SM phase boundary. 

These SM lines are stable in the presence of finite electron--electron interactions $U$ up to a critical $U_c(\alpha,\lambda)$ what is of comparable size as $U_c(1,\lambda)$  (\ie for the isotropic KMH model). Preliminary results for some parameters indicate that for $U>U_c(\alpha\not= 1,\lambda)$ there is a non--magnetic insulator phase like in the $\alpha=1$ case. The detailed investigation of the interacting phase diagram of this plaquette honeycomb model is left for future studies.

\section{Conclusion}
To summarize, through a combination of CDMFT and complementary approaches, we have 
substantiated that the QSH phase is robust toward prominent electron-electron interactions and lattice anisotropies. We have provided a quantitative analysis of the edge state properties which shows that the ordering transition for the spins is also accompanied by a metal-insulator transition at the edges. For very weak SOC, we have
also confirmed the existence of a spin--gapped insulating phase (SL). Finally, we have introduced an anisotropic version of the KM model with the potential to host a rich phase diagram when including the Hubbard interaction.

\begin{acknowledgements}
We acknowledge discussions with F.\,F.\,Assaad, G. Fiete, M. Hohenadler, T. Lang, and G.\,C. Liu. This work was supported by NSF Grant No.\ DMR-0803200 (KLH and SR) and by NSFC under grants Nos.\ 10874235, 10934010, 60978019, by NKBRSFC under grants Nos. 2009CB930701, 2010CB922904, and 2011CB921500 (WW and WML). We acknowledge the Yale
High Performance Computing center and the supercomputing center of CAS for kindly allocating computational resources. 
\end{acknowledgements}

\appendix

\section{CDMFT method}
In order to investigate edge states within CDMFT we use the real--space extension of the homogeneous CDMFT method \cite{kotliar-06rmp865,maier-05rmp1027}. We consider (nano--) ribbons (\ie cylinder geometry) with translation symmetry in direction along the edges (which we aligned for armchair--ribbons along the $y$-axis). 

Within CDMFT we map the original honeycomb lattice onto a $N_c$--site effective cluster embedded in a self-consistent medium. The effective cluster model is obtained via an iterative procedure which can be started with an initial guess of the cluster self--energy $\Sigma({i\omega})$. The effective medium represented by the dynamical mean--field, which is also known as Weiss function $g_c(i\omega)$, is determined by the cluster self--energy $\Sigma(i\omega)$ via the coarse--grained Dyson equation\,\cite{wu-10prb245102}.
Since the translation symmetry in the $x$--direction is broken, we shall use the real--space CDMFT coarse--grained Dyson equation 
for the inhomogeneous system:
\begin{equation}\label{eq:Dyson}
g_{c}^{-1}(i\omega)=\! \Bigg[\sum_{k_{y}}\frac{1}{{i\omega}+\mu-t(k_{y})-\Sigma({i\omega})} \delta_{I,J} \Bigg]^{-1} \!\!\!\! + \Sigma({i\omega})\ .
\end{equation}
Both $g_{c}$ and $\Sigma$ are block diagonal matrices with block size $N_{c}$ and dimension $((N_{c}\times N_{x}) \times (N_{c}\times N_{x}))$. $I, J$ are block indices,  $N_{c}$ is the cluster size, and $N_{x}$ is the number of the clusters after performing the dimensional reduction, \ie $N_x$ corresponds to the {\it height} of the ribbon.

As far as $g_{c}(i\omega)$ is determined, the impurity 
solver\,\cite{Rubtsov} can be used to compute the 
effective many--body impurity problem to obtain the full Green's function $G_{c}$. Eventually, by using 
Dyson equation of the cluster system $\Sigma(i\omega)=g_{c}^{-1}(i\omega)-G_{c}^{-1}(i\omega)$,  
we recalculate the self--energy $\Sigma(i\omega)$  to finish the iterative loop.
The CDMFT self--consistent iterative loop should be repeatedly carried out, until the numerical convergence has been reached. 

Note that we use the spinor notation, $\Psi^{\dagger}=\{ c^{\dag}_{i\up} , c^{\dag}_{i\dw} \} $ ($i\in$ cluster)
 and $\Psi^\pd= (\Psi^{\dagger})^{\dagger}$ since a spontaneous transverse magnetization 
mixes the $\uparrow$ and $\downarrow$ spin--parts of the system.
Therefore the Green's function in the CDMFT framework takes the following form,
\begin{equation}\label{off-diagonalG}\begin{split}
\langle \Psi(\tau_{i}) \Psi^{\dagger}(\tau_{j}) \rangle =  \left[
\begin{array}{ccc}
 G_{\uparrow \uparrow}(\tau_{i}-\tau_{j}) & G_{\uparrow \downarrow}(\tau_{i}-\tau_{j}) \\
 G_{\downarrow \uparrow}(\tau_{i}-\tau_{j}) & G_{\downarrow \downarrow}(\tau_{i}-\tau_{j}) 
\end{array}\right]\ .
\end{split}\end{equation}

 \begin{figure}[b!]
\begin{center}
\includegraphics[scale=0.69]{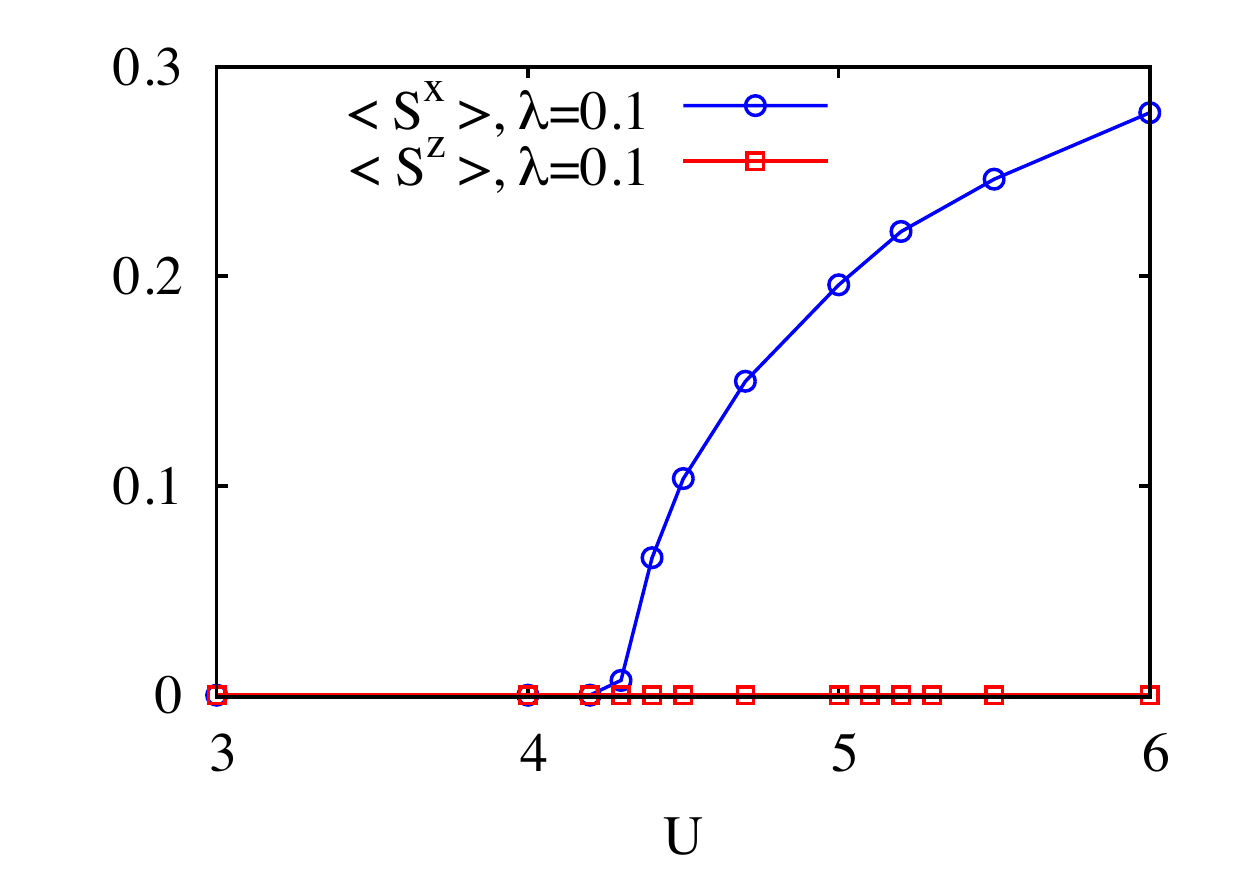}
\caption{(color online). $\langle S^x \rangle$ and $\langle S^z \rangle$ versus $U$ for $\lambda=0.1$.}
\label{fig:sxsz}
\end{center}
\end{figure}

From the above procedure we can easily obtain the quantites $D_{\rm occ}$ \cite{gorelik,wu-10prb245102}
and $\langle S^x \rangle$.
The double occupancy $D_{\rm occ}= \langle n_{i\uparrow}n_{i\downarrow} \rangle= \partial{F} /  \partial{U}$, as the first derivative of the free energy,
is an important quantity in studies of strongly correlated systems. As $D_{occ}$ is directly related to the potential energy this is an indicator for the transition order. 
In the following, we suppress the site index $i$ of the electron operators.
Using the notation of Eq.\,\eqref{off-diagonalG} we find
\begin{equation}
\begin{split}
D_{\rm occ}=&\  \langle c^{\dag}_{\up} c_{\up}^\pd c^{\dag}_{\dw} c_{\dw}^\pd \rangle
=\langle T_{\tau} c^{\dagger}_{\uparrow}(\tau_3) c_{\uparrow} (\tau_2) c^{\dagger}_{\downarrow}(\tau_1) c_{\downarrow}(0) \rangle \\[5pt]
=&\ \langle T_{\tau} c^{\dagger}_{\uparrow}(\tau_3) c_{\uparrow} (\tau_2) \rangle 
  \langle T_{\tau} c^{\dagger}_{\downarrow}(\tau_1) c_{\downarrow}(0) \rangle \\
&\ -\langle T_{\tau} c^{\dagger}_{\uparrow}(\tau_3) c_{\downarrow} (0) \rangle 
  \langle T_{\tau} c^{\dagger}_{\downarrow}(\tau_1) c_{\uparrow} (\tau_2) \rangle \\[5pt]
=&\ G_{\uparrow \uparrow}(0^{-})G_{\downarrow\downarrow}(0^{-}) - G_{\downarrow\uparrow}(0^{-}) G_{\uparrow\downarrow}(0^{+})
\end{split}
\end{equation}
where $0<\tau_1<\tau_2<\tau_3$,
and we used Wick's theorem.

The transverse magnetic order parameter $\langle S_i^x\rangle$ reads 
$\langle S^x_{i}\rangle= \frac{1}{2}\langle c^{\dag}_{i\up} c_{i\dw}^\pd + {\rm h. c.}\rangle$ which can be expressed through the above defined Green's functions,
\begin{equation}
\langle S^x_{i}(\tau=0^-) \rangle= \frac{1}{2}\{ G_{\up\dw}(0^{-})+G_{\dw\up}(0^{-}) \}\ .
\end{equation}
In Fig.\,\ref{fig:sxsz} we show both $\langle S^x \rangle=\langle S_i^x\rangle$ and $\langle S^z\rangle=\langle S_i^z\rangle$ vs.\ $U$ for $\lambda=0.1$. Since the SOC breaks the spin--rotational invariance the magnetization is inside the $XY$--plane; consequently we find a finite value of $\langle S^x\rangle$ at the transition into the SDW phase while $\langle S^z \rangle$ remains zero.

In the case of armchair--ribbons and PBC calculations we used a six--site cluster, for zigzag--ribbons an eight--site cluster, see inset of Fig.\,\ref{fig:phase}. In addition, we performed some PBC calculations with the eight-site cluster as well as with a 24--site cluster (see inset of Fig.\,\ref{fig:aniso}). We found good quantitative agreement between the results performed with the different clusters which underlines the reliability of the CDMFT method. But we would like to emphasize that a six--site cluster is needed to observe the spin liquid phase. For smaller clusters sizes (2 and 4) the spin liquid phase cannot be found as shown in Fig.\,\ref{fig:cluster-size}.


\begin{thebibliography}{10}
\bibitem{spinmodel}
The effective spin Hamiltonian of the KMH model introduced in Ref.\,\onlinecite{rachel-10prb075106} is an anisotropic version of the $J_1$--$J_2$ model which was recently studied in Refs.\,\onlinecite{reuther,mambrini}.
\end{thebibliography}

\end{document}